\documentclass{article}

\usepackage[preprint]{neurips_2025}

\workshoptitle{MATH-AI}

\usepackage[utf8]{inputenc} 
\usepackage[T1]{fontenc}    
\usepackage{hyperref}       
\usepackage{url}            
\usepackage{booktabs}       
\usepackage{amsfonts}       
\usepackage{nicefrac}       
\usepackage{microtype}      
\usepackage{xcolor}         
\usepackage{graphicx}
\usepackage{amsmath}
\usepackage{multirow}
\usepackage{bbding}
\usepackage{bm}
\usepackage{braket}
\usepackage{listings}
\usepackage{textcomp}
\usepackage{makecell}
\usepackage{dsfont}
\usepackage{enumitem}
\usepackage{tablefootnote}

\graphicspath{{./figs/}}
\title{Connected Theorems: A Graph-Based Approach to Evaluating Mathematical Results}

\author{
\textbf{Gergely Bérczi\textsuperscript{1}},
 \textbf{Bin Dong,\textsuperscript{2,3,4}}\thanks{Corresponding author},
  \textbf{Haocheng Ju\textsuperscript{5}},
  \textbf{Tianyi Xu\textsuperscript{6}},
\\
\\
   \textsuperscript{1}Department of Mathematics, Aarhus University\\
\textsuperscript{2}Beijing International Center for Mathematical Research\\ and the New Cornerstone Science Laboratory, Peking University\\
 \textsuperscript{3}Center for Machine Learning Research, Peking University\\
   \textsuperscript{4}Center for Intelligent Computing, Great Bay Institute for Advanced Study, \\Great Bay University\\
 \textsuperscript{5}School of Mathematical Sciences, Peking University\\
 \textsuperscript{6}Beijing International Center for Mathematical Research, Peking University \\
\\
 \texttt{gergely.berczi@math.au.dk}  \quad
 \texttt{dongbin@math.pku.edu.cn} \quad
\\
 \texttt{hcju@pku.edu.cn} \qquad
 \texttt{xutianyi@bicmr.pku.edu.cn} \\
}

\begin{document}

\maketitle

\begin{abstract}
The evaluation of mathematical results plays a central role in assessing researchers' contributions and shaping the direction of the field. Currently, such evaluations rely primarily on human judgment, whether through journal peer review or committees at research institutions. To complement these traditional processes, we propose a data-driven approach. We construct a hierarchical graph linking conjectures, theorems, papers, authors and fields to capture their citation relationships. We then introduce a PageRank-style algorithm to compute influence scores for these entities. Using these scores, we analyze the evolution of field rankings over time and quantify the impact between fields. We hope this framework can contribute to the development of more advanced, quantitative methods for evaluating mathematical research and serve as a complement to expert assessment.

\end{abstract}

\section{Introduction}\label{sec:intro}
Evaluating mathematical results is a critical task, as it influences decisions such as publishing in top journals, recruiting and promoting faculty, assessing grant proposals, and nominating researchers for prestigious prizes. The evaluation of mathematical research is generally guided by two main principles: whether a result supports developments in fields beyond mathematics, and whether it contributes to mathematics itself. In practice, these assessments are conducted by humans through peer review or by panels of distinguished mathematicians.

To complement these traditional evaluation methods and provide a data-driven perspective, we propose a computational approach to quantify the influence of mathematical results from data. We collect published papers that have arXiv preprints from 180 journals and construct a hierarchical five-level graph comprising 6.7k conjectures, 1.88M theorems, 102k papers, and 21 fields. Edges are built based on citation relationships at each level. We then introduce a PageRank-style algorithm that propagates influence both within and across levels to compute influence scores for conjectures, theorems, papers, authors and fields. 

Using these scores, we rank theorems and papers and analyze the evolution of field rankings over time. We observe that PDE \& Dynamical Systems, Geometry, Mathematical Physics, Algebraic Geometry, and Probability emerge as the five most influential fields in recent years. In terms of growth, PDE \& Dynamical Systems, Probability, Mathematical Physics, Numerics, and Optimization appear as the five fastest-growing areas. Examining field-to-field influence further shows that the resulting influence matrix is nearly symmetric, with identifiable clusters such as (Geometry, Analysis I, PDE \& Dynamical Systems) and (PDE \& Dynamical Systems, Probability, Mathematical Physics). Nevertheless, several asymmetries are evident. For example, the influence of Analysis I on PDE is significantly stronger than the reverse, and Mathematical Physics exerts a much greater influence on Numerics than vice versa. These asymmetries reveal directional flows of mathematical ideas, often reflecting the transfer from general theoretical frameworks to more domain-specific applications.

However, as our analysis relies on arXiv versions of published papers, it captures only a subset of the full mathematical literature. In addition, because some citation information is missing or inaccurate, our findings may not fully generalize to the whole mathematical literature. We believe that incorporating additional papers and improving the accuracy of citation extraction at both the theorem and paper levels will refine the five-level graph and further enhance the ability of our approach to complement traditional research evaluation methods.

The rest of the paper is organized as follows. Section \ref{sec:related} reviews the related works on quantitative evaluation of mathematical research and theorem dependency graphs. Section \ref{sec:method} introduces our five-level graph model and the algorithm for computing influence scores. Section \ref{sec:results} presents the experimental results, and Section \ref{sec:conclusion} concludes the paper. Additional details and results are provided in the Appendix.

\section{Related Work}\label{sec:related}

\subsection{Quantitative evaluation of mathematical research}
Citation-based metrics developed for general scientific fields, such as the impact factor for journals and the h-index for authors, can be naively applied to mathematics. However, their use has been criticized in the mathematical context due to their short citation window, limited interpretability, and unstable rankings \cite{1076a2be-10cc-3bdb-8624-1f5569e17b06, 10.1093/reseval/rvv041}. Mathematics-specific measures include the Mathematical Citation Quotient (MCQ), defined as the total citations from MR-indexed journals to articles published in a given journal during the preceding five years, divided by the number of such articles; and the mean cumulative impact factor, a weighted average of rank-normalized impact factors that yields more stable journal rankings \cite{10.1093/reseval/rvv041}. Beyond journals, researcher-level studies report that mathematicians’ output does not decline rapidly with age and that cross-field collaboration correlates with more successful careers \cite{dubois2014productivity}. Work on prize effects finds that Fields Medalists’ productivity (papers per year, citation counts, and student mentoring) declines relative to close contenders after the award, in part because medalists shift to new research areas \cite{borjas2015prizes}. At the subfield level, analyses using indicators such as the number of Fields Medalists, the share of papers from each subfield in top journals, and their presence in top ranked departments suggest that higher status subfields often have a widely recognized list of central open problems \cite{schlenker2020prestige}.

\subsection{Theorem dependency graph}
Theorem dependency graphs can be categorized into two types: formal and natural language. A formal theorem dependency graph is a graph where each node represents a formal definition, axiom, theorem, or lemma, and edges are defined by references among these mathematical objects. Such graphs have been constructed for various formal languages, including Lean \cite{bauer2023mlfmf}, Coq \cite{coq-dpdgraph}, Isabelle \cite{wenzel2014isabelle}, and Agda \cite{bauer2023mlfmf}.

Natural language theorem dependency graphs include NaturalProofs \cite{welleck2021naturalproofs}, AutoMathKG \cite{bian2025automathkg}, and TheoremKB \cite{delemazure:hal-02940819}. Theorems in NaturalProofs are sourced from ProofWiki, the Stacks Project, a real analysis textbook, and a number theory textbook. AutoMathKG \cite{bian2025automathkg} constructed a knowledge graph consisting of definitions, theorems, and problems drawn from ProofWiki, textbooks, a small subset of arXiv papers, and TheoremQA \cite{chen2023theoremqa}, and proposed an automatic update method using large language models.

The most closely related work to ours is TheoremKB \cite{delemazure:hal-02940819}, which constructed a theorem dependency graph with over six million nodes extracted from nearly all mathematics papers on arXiv. Their work differs from ours in several respects. First, their dataset includes almost all mathematics papers on arXiv, many of which are not peer reviewed. In contrast, we restrict our attention to published papers and use their arXiv versions, as peer-reviewed papers are generally regarded as higher-quality research outputs. Second, TheoremKB focuses on building a theorem dependency graph, whereas we develop a five-level graph that integrates conjectures, theorems, papers, authors and fields, enabling analysis at multiple levels of abstraction. Third, their work concentrates on the technical aspects of graph construction and provides limited analysis, mainly concerning the largest connected component and the most cited results. By comparison, our study not only details the model, algorithms, and data construction process, but also investigates the evolution of field rankings and the field-to-field impact.

\section{Methodology}\label{sec:method}
In this section, we describe our five-level graph that models the citation relationships among conjectures, theorems, papers, authors and fields. After introducing the model, we present our algorithm for computing the influences of conjectures, theorems, papers, authors and fields.

\subsection{Constructing the five-level graph}

\begin{figure}[t]
\centering
\includegraphics[width=0.4\linewidth]{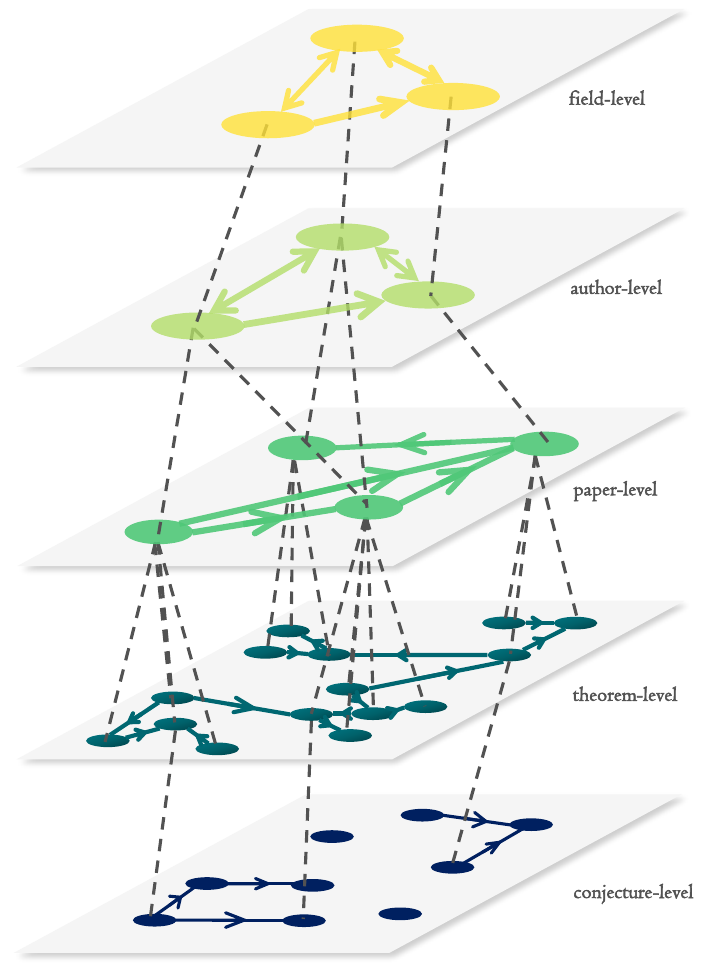}
\caption{The five-level graph.}
\label{fig:5-level-graph}
\end{figure}

\begin{table}[h]
\caption{We follow the classification used in \cite{schlenker2020prestige}. The left column lists mathematical fields, and the right column shows the corresponding two-digit Mathematics Subject Classification (MSC) codes grouped under each field.}
\label{tab:msc_code}
\centering
\resizebox{0.99\linewidth}{!}{
\begin{tabular}{lc}
\toprule
Math field & 2-digit MSC code\\
\midrule
Logic \& Foundations & 03 \\
Combinatorics \& Discrete Mathematics & 05, 06 \\
Algebra (general, rings, categories, etc.) & 08, 15, 16, 18 \\
Lie Theory \& Representation / Topological Groups & 17, 20, 22 \\
Number Theory \& Arithmetic & 11, 12 \\
Algebraic Geometry \& Commutative Algebra & 13, 14 \\
Geometry (differential/convex/discrete) & 51, 52, 53, 58 \\
Topology \& Geometry of Manifolds & 19, 54, 55, 57 \\
Analysis I (Real/Complex/Harmonic) & 26, 28, 30, 31, 32, 33, 40, 41, 42, 43 \\
Analysis II (Functional Analysis \& Operators) & 46, 47 \\
Differential Equations \& Dynamical Systems (ODE/PDE) & 34, 35, 37, 39, 44, 45, 49 \\
Probability \& Stochastic Processes & 60 \\
Statistics & 62 \\
Numerical Analysis \& Scientific Computing & 65, 68 \\
Optimization \& Operations Research & 90 \\
Mathematical Physics \& Mechanics & 70, 74, 76, 78, 80, 81, 82, 83, 85, 86 \\
Systems \& Control / Information Theory & 93, 94 \\
Economics \& Social Sciences / Game Theory & 91 \\
Biology \& Other Natural Sciences & 92 \\
General / History / Education / Misc. & 00, 01, 97 \\
Others & other two-digit MSC codes \\
\bottomrule
\end{tabular}}
\end{table}

\begin{figure}[tb]
\centering
\includegraphics[width=0.99\linewidth]{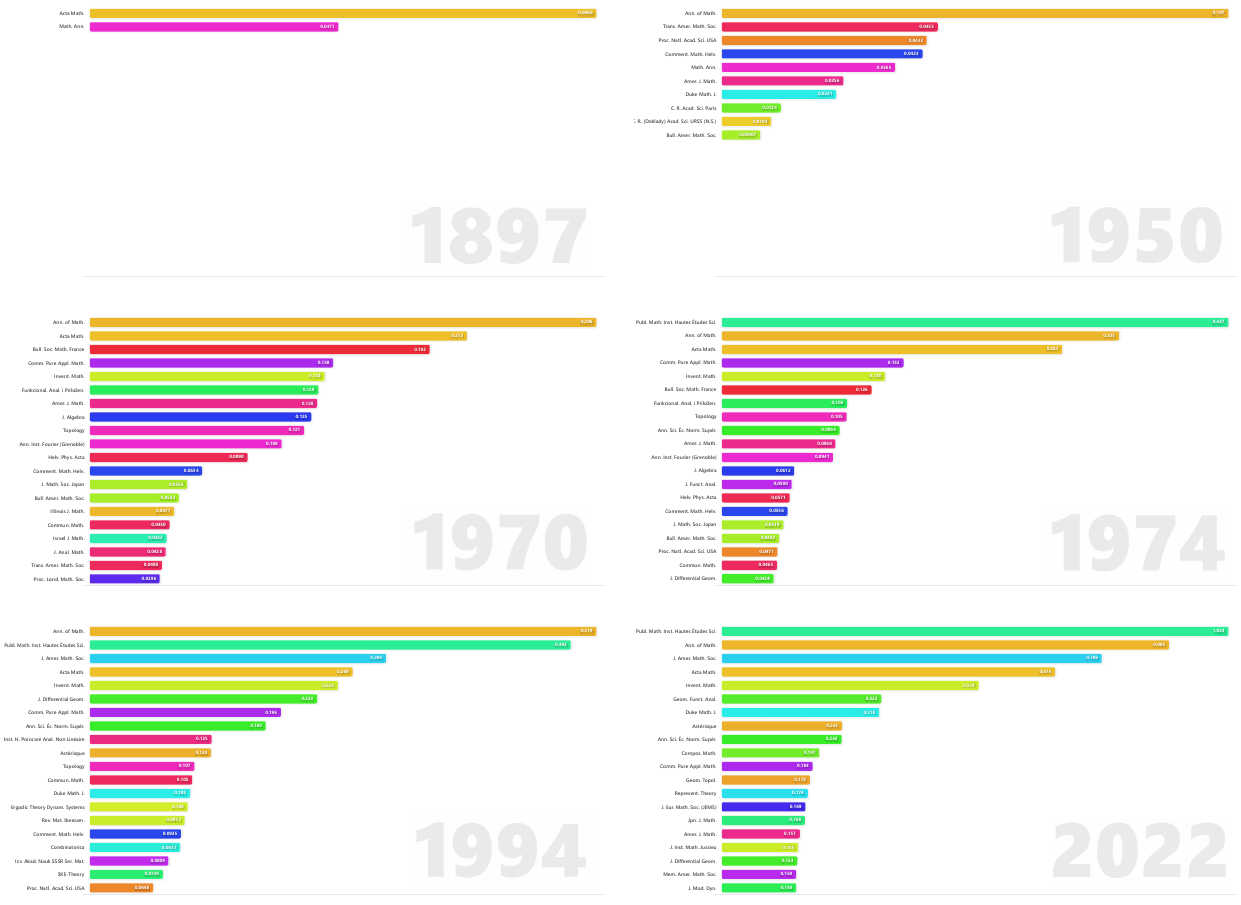}
\caption{Journal rankings for the years 1897, 1950, 1970, 1974, 1994, and 2022. In 1897, only two journals appear in the ranking; this does not imply that only two journals existed at that time. Rather, we include only journals that received more than 50 ICM citations and published more than 100 papers in the 15 years preceding each respective year.}
\label{fig:journal_rank}
\end{figure}

We begin by selecting 180 journals using the following procedure. First, we collect all ICM proceedings throughout history and extracted the references cited in each proceeding paper. For each journal, we then compute the total number of citations it received from ICM proceedings to papers published between 2007 and 2021. We retain only journals that received more than 50 such citations and published more than 100 papers during 2007-2021. The complete list of selected journals is provided in Appendix \ref{sec:app_journals}.

For each ICM year, we further rank journals using a score defined as follows: the total number of citations a journal receives from ICM proceedings to papers published in the preceding 15 years, divided by the number of papers it publishes during the same period. The top journals in six representative years according to this metric are shown in Figure \ref{fig:journal_rank}. In 1897, Acta Mathematica ranks first, despite having been founded only in 1882. In 1950, Annals of Mathematics ranks first. In 1970, Inventiones Mathematicae enters the top five, only four years after its founding in 1966. In 1974, Publications Mathématiques de l'IHÉS ranks first, having been established in 1959. In 1994, Journal of the American Mathematical Society enters the top three, just six years after its founding in 1988. These examples suggest that leading journals often achieve top rankings shortly after their establishment. This rapid rise may be attributed to their strong founding leadership, early academic prestige, and the high quality of initial submissions.

We collect, from 180 journals, published papers that have arXiv preprints. To categorize papers into their primary fields, we use the first two digits of the Mathematics Subject Classification (MSC) code. The classification of these fields is shown in Table \ref{tab:msc_code}.

We then construct a five-level directed graph $\mathcal{G} = (\mathcal{C}, \mathcal{T}, \mathcal{P}, \mathcal{A}, \mathcal{F}, \mathcal{E})$ where $\mathcal{C}$ is the set of conjectures, $\mathcal{T}$ is the set of theorems, $\mathcal{P}$ is the set of papers, $\mathcal{A}$ is the set of authors, $\mathcal{F}$ is the set of mathematical fields, and $\mathcal{E}$ is the set of edges. As shown in Figure \ref{fig:5-level-graph}, the graph is organized hierarchically into five levels: the conjecture level, the theorem level, the paper level, the author level and the field level. The intra-level edges are defined as follows:
\begin{itemize}
    \item Conjecture level: There is an edge from conjecture $c_i$ to conjecture $c_j$ if $c_j$ implies $c_i$.
    \item Theorem level: There is an edge from theorem $t_i$ to theorem $t_j$ if the proof of $t_i$ cites $t_j$.
    \item Paper level: There is an edge from paper $p_i$ to paper $p_j$ if $p_i$ cites $p_j$.
    \item Author level: There is an edge from author $a_i$ to author $a_j$ if $a_i$'s paper cites $a_j$'s paper, or if $a_i$ and $a_j$ are co-authors.
    \item Field level: There is an edge from field $f_i$ to field $f_j$ if at least one paper in $f_i$ cites a paper in $f_j$.
\end{itemize}
The cross-level edges are defined as follows:
\begin{itemize}
    \item For each conjecture $c \in \mathcal{C}$, there is a bidirectional edge between $c$ and the theorem $t$ that attempts to resolve it.
    \item For each theorem $t \in \mathcal{T}$, there is a bidirectional edge between $t$ and the paper it belongs to.
    \item For each paper $p \in \mathcal{P}$, there is a bidirectional edge between $p$ and its authors.
    \item For each author $a\in \mathcal{A}$, there is a bidirectional edge between $a$ and their field(s).
\end{itemize} 

We denote the following neighbor sets:

\[
\begin{array}{ll}
N_C(c) = \{c' \in \mathcal{C} \mid c'\to c\},  \quad N_T(t) = \{t' \in \mathcal{T} \mid t'\to t\},\quad
N_P(p) = \{p' \in \mathcal{P} \mid p'\to p \}, \\
N_A(a) = \{a' \in \mathcal{A} \mid a'\to a \},\quad
N_F(f) = \{f' \in \mathcal{F} \mid f'\to f \}.
\end{array}
\]
We define the following mappings:
\begin{itemize}
    \item $\phi_{TC}(t)$: the conjecture that theorem $t$ attempts to resolve;
    \item $\phi_{TP}(t)$: the paper that theorem $t$ belongs to;
    \item $\phi_{PF}(p)$: the primary field of paper $p$;
    \item $\phi_{PT}(p) = \{t \in \mathcal{T} \mid t \text{ is in } p\}$: the set of theorems in paper $p$;
    \item $\phi_{FP}(f) = \{p \in \mathcal{P} \mid p \text{ belongs to } f\}$: the set of papers in field $f$;
    \item $\phi_{AP}(a)=\{p\in \mathcal{P} \mid a \text{ is one of the authors of } p\}$: the set of papers belonging to $a$;
    \item $\phi_{PA}(p)=\{a\in \mathcal{A} \mid a \text{ is one of the authors of } p\}$: the authors of paper $p$;
    \item $\phi_{AF}(a)$: the primary field of author $a$;
    \item $\phi_{FA}(f)=\{a\in \mathcal{A} \mid \phi_{AF}(a)= f\}$: the set of authors whose primary field is $f$.
\end{itemize}

The weight matrices for each level are defined as follows. For the conjecture level:
\[
C(c_i,c_j) = 
\left\{
\begin{array}{ll}
  0   & c_j \notin N_C(c_i) \text{ and } c_i, c_j \text{ are unrelated}, \\
  0.5 & c_j \notin N_C(c_i), \text{but }c_j \text{ is related to } c_i,      \\
  1 & c_j \in N_C(c_i).
\end{array}
\right.
\]

For the theorem level:

 \[
T(t_i,t_j) = 
\left\{
\begin{array}{ll}
0 &  t_j \notin N_T(t_i), \\
0.05 &  t_j \in N_T(t_i) \text{ and } t_i, t_j \text{ are from the same paper,} \\
0.1  &  t_j \in N_T(t_i) \text{ and } t_i, t_j \text{ are from different papers, but share at least one author,} \\
1    & t_j \in N_T(t_i) \text{ and } t_i, t_j \text{ are from different papers with no shared authors.}
\end{array}
\right.
\]
For the paper level:
\[
P(p_i,p_j) = 
\left\{
\begin{array}{ll}
  0   & p_j \notin N_P(p_i), \\
  0.1   &  p_j \in N_P(p_i) \text{ and } p_i, p_j \text{ share at least one author,} \\
  1 & p_j \in N_P(p_i) \text{ and } p_i, p_j \text{ have no shared authors.}
\end{array}
\right.
\]

For the author level: we define $A(a_i,a_j)$ as the number of citations from $a_j$'s papers to $a_i$'s papers (excluding citations between their co-authored papers), plus the number of papers co-authored by $a_i$ and $a_j$.
\[
A(a_i,a_j)= \sum_{p_1\in \phi_{AP}(a_i)\backslash \phi_{AP}(a_j),p_2\in\phi_{AP}(a_j)\backslash \phi_{AP}(a_i)} \mathds{1}_{\{P(p_1,p_2)>0\}} + \lvert \phi_{AP}(a_i)\cap \phi_{AP}(a_j) \rvert
\]

For the field level:
\[
F(f_i,f_j) = \sum_{p_1\in f_i, p_2\in f_j }\mathds{1}_{\{P(p_1,p_2)>0\}}.
\]

We have not yet conducted a sensitivity analysis to evaluate the robustness of field rankings across a reasonable range of hyperparameters; this will be addressed following feedback from domain experts.

\subsection{Computing the influences}
Let $u_C(c), u_T(t)$, $u_P(p)$, $u_A(a)$ and $u_F(f)$ denote the influence scores of conjecture $c$, theorem $t$, paper $p$, author $a$, and field $f$, respectively. Let $\widetilde{C}, \widetilde{T}, \widetilde{P}, \widetilde{A}, \widetilde{F}$ be the column-normalized weight matrices:
\[
\begin{array}{ll}
&\widetilde{C}(c_i,c_j) = \frac{C(c_i,c_j)}{\sum_k C(c_k,c_j)}, \qquad  \widetilde{T}(t_i,t_j) = \frac{T(t_i,t_j)}{\sum_k T(t_k,t_j)}, \qquad \widetilde{P}(p_i,p_j) = \frac{P(p_i,p_j)}{\sum_k P(p_k,p_j)},\\
&\widetilde{A}(a_i,a_j) = \frac{A(a_i,a_j)}{\sum_k A(a_k,a_j)}, \qquad \widetilde{F}(f_i,f_j) = \frac{F(f_i,f_j)}{\sum_k F(f_k,f_j)}.
\end{array}
\]

We aim to propagate influence across the five levels. At the conjecture level, each conjecture accumulates influence from the conjectures that imply it.
At the theorem level, each theorem accumulates influence from the theorems that cite it, inherits influence from its corresponding paper, and receives additional score if it is cited in an ICM proceeding or is related to resolving a conjecture.
At the paper level, each paper accumulates influence from citing papers, inherits influence from its authors, incorporates the influence of its most influential theorem, and receives additional score if it is published in a top-20 journal or cited in ICM proceedings.
At the author level, each author accumulates influence from citing authors, inherits influence from their primary field, incorporates the influence of their most influential paper, and receives additional score if the author is an ICM speaker. At the field level, each field aggregates influence from citing fields and from the authors within the field whose influence exceeds the average value $\frac{1}{|\mathcal{A}|}$. Based on this structure, we propose an iterative update of $u_C(c), u_T(t)$, $u_P(p)$, $u_A(a)$ and $u_F(f)$ following the PageRank algorithm:
\[
\left\{
\begin{array}{ll}
\hat{u}^{(k+1)}_C(c) =& \sum_{c'\in N_C(c)}\widetilde{C}(c,c')u^{(k)}_C(c') \\
\hat{u}^{(k+1)}_T(t) =& \alpha_T\sum_{t'\in N_T(t)}\widetilde{T}(t,t')u^{(k)}_T(t')+\beta_T\frac{u^{(k)}_P(\phi_{TP}(t))}{|\mathcal{T}|/|\mathcal{P}|}+\gamma_T u_{ICM_t}(t)\\
&+(1-\alpha_T-\beta_T-\gamma_T)\mathds{1}_{\{t\in \mathcal{T}_c\}}u_C^{(k)}(\phi_{TC}(t))  \\
\hat{u}^{(k+1)}_P(p) =& \alpha_P\sum_{p'\in N_P(p)}\widetilde{P}(p,p') u^{(k)}_P(p')+\beta_P \frac{\frac{1}{\lvert\phi_{PA}(p)\rvert }\sum_{a'\in \phi_{PA}(p)}u^{(k)}_A(a')}{|\mathcal{P}|/|\mathcal{A}|} \\
&+ \gamma_P\max_{t\in\phi_{PT}(p)} u_T^{(k)}(t)  +\delta_p u_J(p) + (1-\alpha_P-\beta_P-\gamma_P-\delta_p)u_{ICM_p}(p) \\
\hat{u}^{(k+1)}_A(a) =& \alpha_A\sum_{a'\in N_A(a)}\widetilde{A}(a,a') u^{(k)}_A(a')+\beta_A \frac{u^{(k)}_F(\phi_{AF}(a))}{|\mathcal{A}|/|\mathcal{F}|}\\
&+\gamma_A\max_{p\in\phi_{AP}(a)} u_P^{(k)}(p)+(1-\alpha_A-\beta_A-\gamma_A)u_{ICM_a}(a)\\
\hat{u}^{(k+1)}_F(f) =& \alpha_F\sum_{f'\in N_F(f)}\widetilde{F}(f,f') u^{(k)}_F(f')+(1-\alpha_F)\sum_{a\in\phi_{FA}(f)} \left(u_A^{(k)}(a) - \frac{1}{|\mathcal{A}|}\right)_+
\end{array}
\right.
\]

where $\hat{u}^{(k+1)}_C(c),\hat{u}^{(k+1)}_T(t), \hat{u}^{(k+1)}_P(p), \hat{u}^{(k+1)}_A(a), \hat{u}^{(k+1)}_F(f)$ are the unnormalized scores at iteration $k+1$.

\begin{itemize}
    \item $u_{ICM_t}(t)=\frac{1}{|\mathcal{T}|}$ if the theorem is cited by ICM proceedings, otherwise $u_{ICM_t}(t)=0$.
    \item $\mathcal{T}_c$ is the set of theorems that attempt to resolve a conjecture.
    \item $u_J(p)=\frac{v_J(\phi_{PJ}(p))}{|\mathcal{P}|}$, where $\phi_{PJ}(p)$ is the journal of paper $p$, $v_J(j)$ is the weight for journal $j$.

   \[
v(j)=
\begin{cases}
1, & rank(j)=1,\dots,6,\\
\exp\left(-\dfrac{rank(j)-6}{2}\right), & rank(j)=7,\dots,20,\\
0, & other\ journals.
\end{cases}
\]
    \item $u_{ICM_p}(p)=\frac{1}{|\mathcal{P}|}$ for papers cited by ICM proceedings and $u_{ICM_p}(p)=0$ for others.
    \item $u_{ICM_a}(a)=\frac{2}{|\mathcal{A}|}$ for plenary speakers, $u_{ICM_a}(a)=\frac{1}{|\mathcal{A}|}$ for other ICM speakers, $u_{ICM_a}(a)=0$ for other mathematicians.
    \item $(x)_+ = \max(x, 0)$.
\end{itemize}

We then normalize the updated scores to unit $\ell_1$ norm:
\[
\begin{array}{ll}
&u^{(k+1)}_C(c)=\frac{\hat{u}^{(k+1)}_C(c)}{\sum_{c'\in\mathcal{C}}\hat{u}^{(k+1)}_C(c')},
u^{(k+1)}_T(t)=\frac{\hat{u}^{(k+1)}_T(t)}{\sum_{t'\in\mathcal{T}}\hat{u}^{(k+1)}_T(t')}, u^{(k+1)}_P(p)=\frac{\hat{u}^{(k+1)}_P(p)}{\sum_{p'\in\mathcal{P}}\hat{u}^{(k+1)}_P(p')},\\
&u^{(k+1)}_A(a)=\frac{\hat{u}^{(k+1)}_A(a)}{\sum_{a'\in\mathcal{A}}\hat{u}^{(k+1)}_A(a')},
u^{(k+1)}_F(f)=\frac{\hat{u}^{(k+1)}_F(f)}{\sum_{f'\in\mathcal{F}}\hat{u}^{(k+1)}_F(f')}.
\end{array}
\]
Here, $u_C^{(k+1)}(c)$, $u_T^{(k+1)}(t)$, $u_P^{(k+1)}(p)$, $u_A^{(k+1)}(a)$, and $u_F^{(k+1)}(f)$ denote the normalized influence scores at iteration $k+1$, and $\alpha_T, \beta_T, \gamma_T, \alpha_P,\beta_P,\gamma_P, \delta_P, \alpha_A, \beta_A, \gamma_A, \alpha_F \in (0,1)$ with $\alpha_T+\beta_T+\gamma_T<1, \alpha_P + \beta_P+\gamma_P+\delta_P < 1, \alpha_A + \beta_A +\gamma_A< 1$. The scores are initialized uniformly as $u^{(0)}_C(c)=\frac{1}{|\mathcal{C}|},u^{(0)}_T(t)=\frac{1}{|\mathcal{T}|}, u^{(0)}_P(p)=\frac{1}{|\mathcal{P}|}, u^{(0)}_A(a)=\frac{1}{|\mathcal{A}|}, u^{(0)}_F(f)=\frac{1}{|\mathcal{F}|}.$ The stopping criterion is
\[
\begin{array}{ll}
\max\{&\sum_{c\in\mathcal{C}}|u^{(k+1)}_C(c)-u^{(k)}_C(c)|,\sum_{t\in\mathcal{T}}|u^{(k+1)}_T(t)-u^{(k)}_T(t)|, \sum_{p\in\mathcal{P}}|u^{(k+1)}_P(p)-u^{(k)}_P(p)|, \\
&\sum_{a\in\mathcal{A}}|u^{(k+1)}_A(a)-u^{(k)}_A(a)|,
\sum_{f\in\mathcal{F}}|u^{(k+1)}_F(f)-u^{(k)}_F(f)|\}<10^{-7}.
\end{array}
\]

\section{Results}\label{sec:results}
In this section, we examine macro-level characteristics of mathematical fields, including the temporal evolution of field-level influence scores and inter-field impact.

\subsection{Evaluating field influences over time}

For each year from 2010 to 2023, we construct the five-level graph by including all papers whose first arXiv version appeared before the end of that year. For example, the graph for the year 2010 includes papers published between 1991 and December 2010. 

\begin{figure}[tb]
\centering
\includegraphics[width=0.99\linewidth]{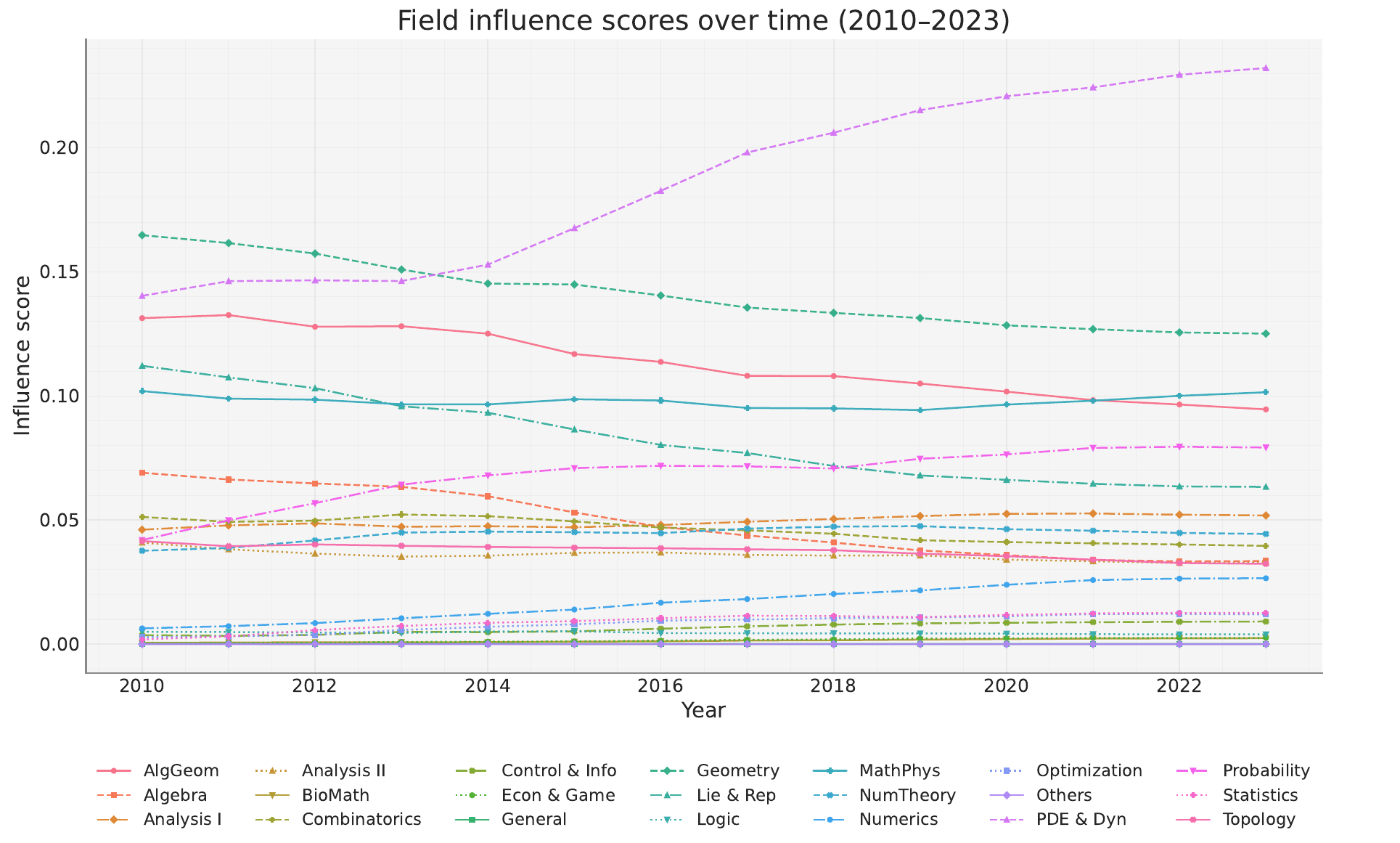}
\caption{Influence scores of fields over time. $(\alpha_T, \beta_T, \gamma_T, \alpha_P, \beta_P, \gamma_P, \delta_P, \alpha_A, \beta_A, \gamma_A, \alpha_F) = (0.8, 0.1, 0.05, 0.6, 0.2, 0.1, 0.05, 0.6, 0.2, 0.1, 0.85)$.}
\label{fig:field-score}
\end{figure}

\begin{figure}[tb]
\centering
\includegraphics[width=0.8\linewidth]{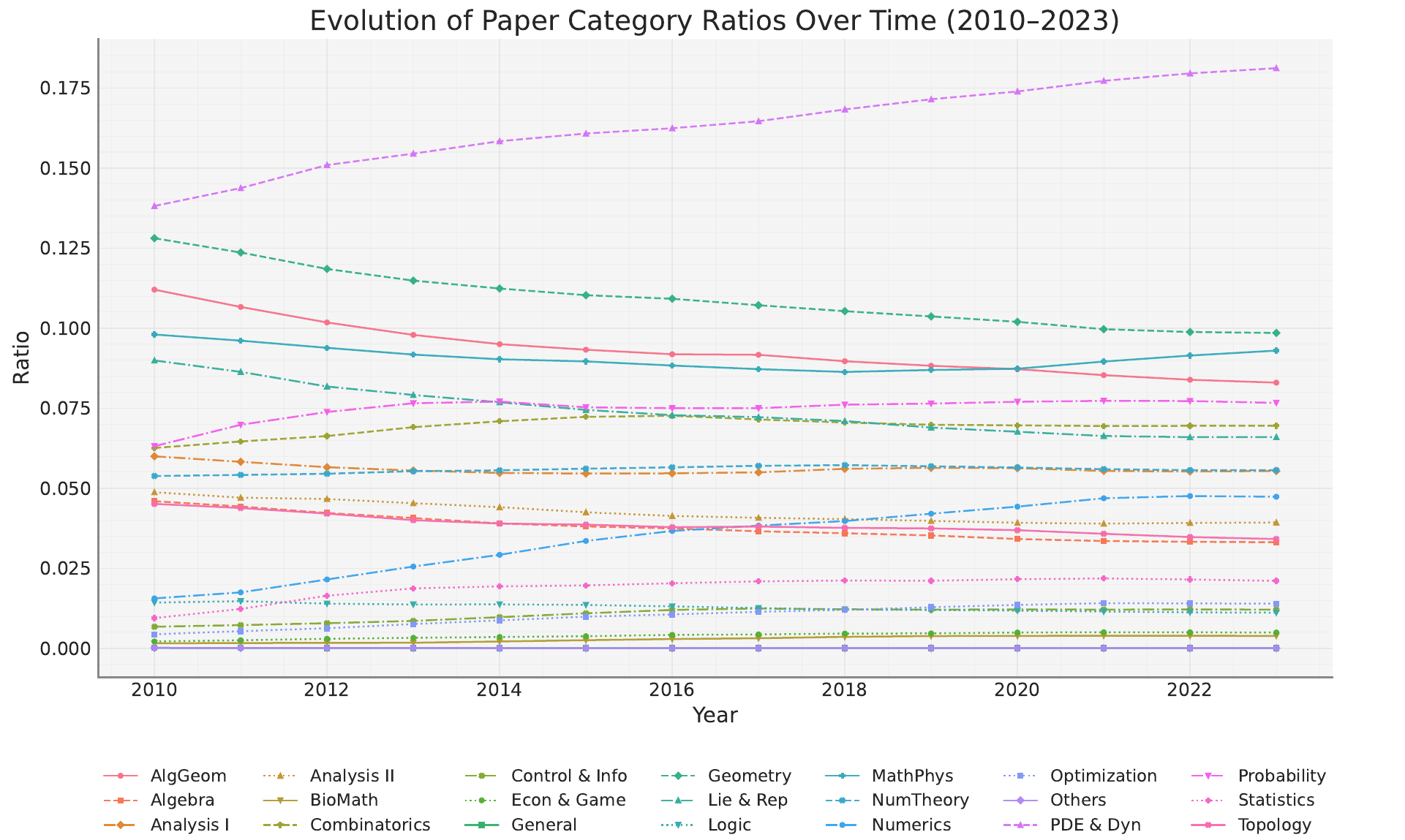}
\caption{Evolution of paper category ratios over time. For each year, the ratio for each category is calculated by dividing the cumulative number of papers in that category from 1991 up to the end of that year by the total number of papers across all categories during the same period.}
\label{fig:ratio}
\end{figure}

We illustrate the results under the hyperparameter setting $(\alpha_T, \beta_T, \gamma_T, \alpha_P, \beta_P, \gamma_P, \delta_P, \alpha_A, \beta_A, \gamma_A,\\ \alpha_F) = (0.8, 0.1, 0.05, 0.6, 0.2, 0.1, 0.05, 0.6, 0.2, 0.1, 0.85)$ in Figure \ref{fig:field-score}. In recent years, PDE \& Dyn, Geometry, MathPhys, AlgeGeom, and Probability are the top-5 fields. PDE \& Dyn and Probability exhibit steady growth over the years. It is worth noting that although the proportion of Probability papers did not increase after 2013 and even declined slightly (see Figure \ref{fig:ratio}), their influence score continues to rise.

\subsection{Evaluating the field-to-field impact}
\begin{figure}[tb]
\centering
\includegraphics[width=0.95\linewidth]{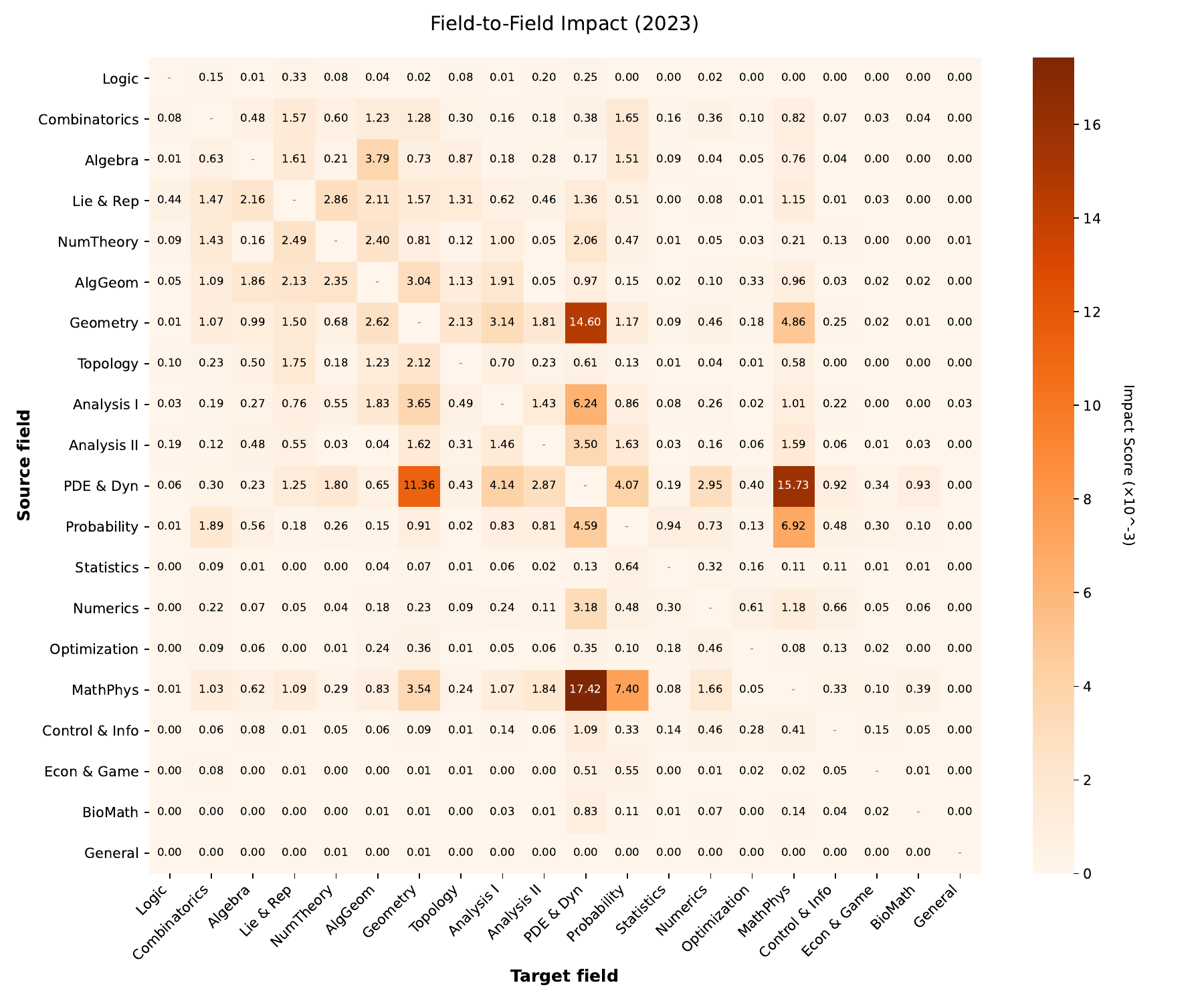}
\caption{Field-to-field impact.}
\label{fig:f2f-impact}
\end{figure}

We aim to quantify the total influence that field $f$ exerts on field $f'$ through paper-level citations. The underlying idea is that the more influential a citing paper $p'$ is (as reflected by a large $u_P(p')$), the greater impact it assigns to the papers it cites. Accordingly, we define the impact of field $f$ on field $f'$  as
\[
I(f,f')=\sum_{p\in f, p'\in f'}\widetilde{P}(p,p')u_P(p').
\]
The results are shown in Figure \ref{fig:f2f-impact}, with hyperparameters set to $(\alpha_T, \alpha_P, \beta_P, \alpha_F) = (0.6, 0.6, 0.05, 0.85)$ and data up to December 2023. The heatmap reveals several patterns:

\paragraph{Near-Symmetric Field Pairs.} Many field pairs have nearly symmetric influence, indicating balanced knowledge exchange. Examples include (Geometry, Topology), (NumTheory, AlgGeom).

\paragraph{Clusters of Fields.} The heatmap also reveals field clusters in which each pair exhibits strong mutual influence, such as  (Geometry, Analysis I, PDE \& Dyn). In addition, the (PDE \& Dyn, Probability, MathPhys)  exhibits consistently high impact scores in both directions, reflecting their close interconnection.

\paragraph{Asymmetric Relationships.} Despite overall symmetry, several pairs display imbalance. For example, the impact of Analysis I on PDE is 51\% higher than the reverse. The stronger influence of Analysis on PDE is natural, since many foundational techniques in PDE originate from Analysis, whereas feedback from PDE to Analysis is comparatively weaker. Similarly, the impact of Mathematical Physics on Numerics is 41\% greater than the reverse. This asymmetry likely reflects the fact that numerical analysis and scientific computing primarily address problems arising from mathematical physics, while comparatively fewer conceptual advances flow back in the opposite direction.

\begin{table}[tb]
\caption{Top-5 fastest-growing research fields over the past five years.}
\label{tab:fast_grow_fields}
\centering
\begin{tabular}{lc}
\toprule
Math field & Average Growth Rate\\
\midrule
PDE \& Dyn & 0.0052 \\
Probability&  0.0017\\
MathPhys  &  0.0013 \\
Numerics  &  0.0013\\
Optimization   &     0.0003 \\
\bottomrule
\end{tabular}
\end{table}

\section{Conclusion}\label{sec:conclusion}
In this section, we propose a computational framework for studying the temporal development of mathematical fields and the influence they exert on one another. We collect published papers that have arXiv preprints and construct a five-level graph comprising conjectures, theorems, papers, authors and fields. Intra-level edges are defined by citation relationships, while inter-level edges capture hierarchical associations: theorems are linked to the papers they appear in, and papers are linked to their corresponding fields. To compute influence scores across all levels, we develop a PageRank-style algorithm that propagates influence both within and across levels. This allows us to assign influence scores to conjectures, theorems, papers, authors and fields, and to rank entities accordingly. Using these scores, we analyze the evolution of field influence over time. Our results show that PDE and Dynamical Systems, Geometry, Mathematical Physics, Algebraic Geometry, and Probability are the five most influential fields in recent years. In terms of growth, PDE and Dynamical Systems, Probability, Mathematical Physics, Numerics, and Optimization emerge as the five fastest-growing fields. Furthermore, when examining field-to-field impact, we find that the resulting influence matrix is nearly symmetric, with identifiable clusters such as (Geometry, Analysis I, PDE \& Dynamical Systems) and (PDE \& Dynamical Systems, Probability, Mathematical Physics). Nevertheless, certain asymmetries are apparent. For instance, the impact of Analysis I on PDE is significantly stronger than the reverse. These patterns suggest a directional flow of knowledge across mathematical domains: fields that develop general theoretical frameworks tend to exert greater influence on more application-driven areas.

\section*{Limitations}
Despite these findings, several limitations of the current framework remain and point to directions for future work. First, the influence propagation algorithm could be refined to better reflect mathematical practice. Expert knowledge could be incorporated to guide the weighting scheme, for instance by assigning higher weights to papers published in leading journals within each subfield. Second, the accuracy of citation information can be further improved. In practice, a theorem may be used in a proof without being explicitly cited at the theorem level; references may appear only in the introduction, or authors may cite a paper without naming the specific result they rely on. These behaviors lead to missing or incomplete theorem-level citations. Future work could develop more refined extraction methods to surface such implicit dependencies. In addition, some nodes may need to be merged. Certain theorems are essentially restatements of the same result and should therefore be consolidated into a single node. For the conjecture layer, we compare each conjecture with its top-30 most similar conjectures to detect duplicates. However, the theorem layer is much larger than the conjecture layer, and more efficient methods will be required to identify and merge duplicate theorem nodes. Third, the corpus itself can be expanded to provide broader coverage of the mathematical literature, especially papers published before the advent of arXiv as well as textbooks, in order to more fully capture the historical and structural landscape of mathematics.

\begin{ack}
This work is supported in part by the National Key R\&D Program of China grant 2024YFA1014000, the Fundamental and Interdisciplinary Disciplines Breakthrough Plan of the Ministry of Education of China (JYB2025XDXM113), and the New Cornerstone Investigator Program.
\end{ack}

\bibliographystyle{plain}
\bibliography{ref}


\appendix

\newpage
\section{Details of Data Collection and Graph Construction}\label{sec:app_details}
In this section, we describe how we collect the papers, extract citation information at the conjecture, theorem and paper levels, and construct the five-level graph.

\subsection{Data Collection}

\begin{figure}[tb]
\centering
\includegraphics[width=0.95\linewidth]{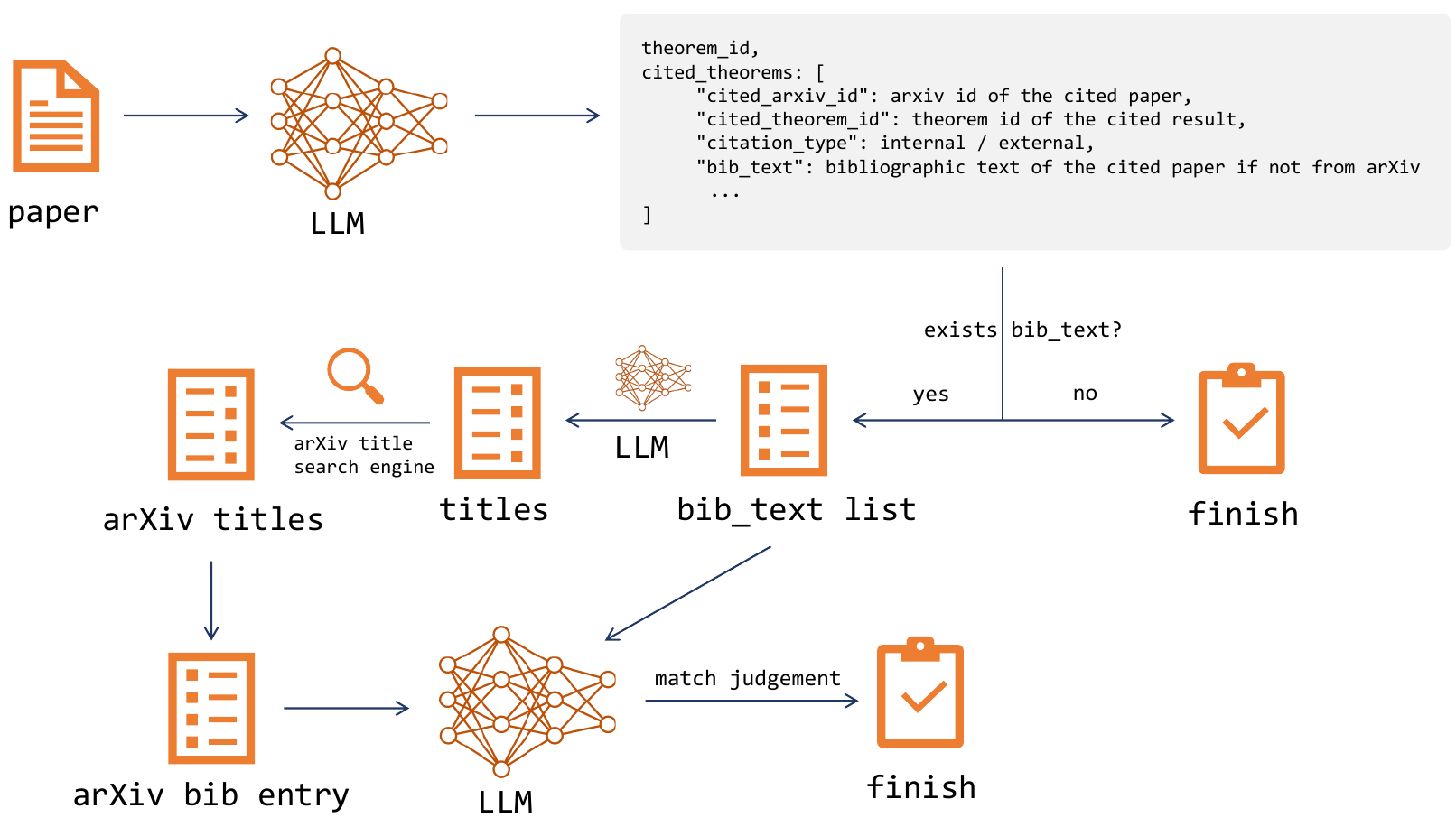}
\caption{Our agent for extracting theorem-level citations and matching published papers to arXiv papers. We first use a large language model to extract theorems from a paper together with the theorems cited in their proofs. If all cited theorems originate from arXiv papers, the process terminates. Otherwise, for cited theorems coming from published papers, we determine whether they correspond to published versions of arXiv papers. This is done by extracting the title from the bibliographic text, matching it against arXiv titles to obtain the most similar candidate, and then using a large language model to judge whether the bibliographic entry matches the arXiv metadata. Only bibliographic entries that pass this verification are treated as having corresponding arXiv papers.}
\label{fig:cnt-agent}
\end{figure}

We begin by selecting 180 journals according to the procedure described in Section \ref{sec:method}. From the ICM proceedings throughout history, we extract conjectures mentioned in the papers and use them to construct the conjecture layer. Next, using the Semantic Scholar Academic Graph\footnote{\href{https://api.semanticscholar.org/api-docs/datasets}{https://api.semanticscholar.org/api-docs/datasets}} \cite{10.1145/3487553.3527147}, we then collect all papers that have an arXiv version and keep only those with a DOI. Each DOI is used to query MathSciNet in order to obtain the MSC code and author IDs, which allow us to categorize each paper by its primary field and to determine author overlap when computing edge weights. This process yields tuples of the form (arXiv ID, journal, category, author IDs).

\subsection{Graph Construction}

The graph consists of five types of nodes: conjectures, theorems, papers, authors and fields. Since author-level and field-level edges are derived from paper-level citations, our construction primarily focuses on establishing edges at the paper, theorem, and conjecture levels. For the paper-level edges, we rely on the Semantic Scholar Academic Graph \cite{10.1145/3487553.3527147} to obtain the list of cited papers for each paper.

To construct theorem-level edges, we design an agent that extracts theorem-level citations and aligns them with published papers. The overall workflow of the agent is illustrated in Figure \ref{fig:cnt-agent}. First, we use a large language model (LLM) to process the full text of each paper and extract all theorems appearing in the paper, together with the theorems referenced in the proof of each theorem. In our experiments, we use DeepSeek-V3.1-Terminus for this task. The model produces a structured list in which each element corresponds to a theorem and contains its theorem identifier, along with a list of theorems used in its proof. Each cited theorem in this list is represented by the following fields: the arXiv ID of the paper to which the theorem belongs. If the theorem comes from the current paper, we use the current paper's arXiv ID. If it comes from a published paper or a book, this field is left empty. The second field is the cited theorem's identifier. The third field records the citation type, which is internal when the cited theorem is from the same paper and external otherwise, since different weights are assigned to internal and external citations. The final field contains the bibliographic text of the cited source. This bibliographic text is left empty when the cited theorem originates from the current paper or another arXiv paper, and is provided only for theorems cited from published papers or books. This bibliographic information is later used to match cited results to arXiv papers.

If all bibliographic text fields are empty, then all cited theorems originate from arXiv papers, and we can directly construct theorem-level edges between arXiv theorems in our dataset. If some bibliographic text fields are non-empty, this indicates that certain cited theorems come from published papers or books, and we attempt to determine whether these correspond to published versions of arXiv papers in our corpus. To do so, we first use an LLM to extract the title from each bibliographic entry. Each extracted title is then used to query an arXiv title semantic search engine that we have built. This search engine uses all arXiv paper titles as its corpus and represents them in a vector database constructed using OpenAI's \texttt{text-embedding-ada-002} model. Query titles are embedded using the same model, and cosine similarity is used to retrieve the most similar arXiv titles. If the top-1 similarity score exceeds 0.97, we treat the closest match as a candidate arXiv paper; otherwise, the entry is discarded.

For retained candidates, we further use an LLM to verify whether the bibliographic entry matches the arXiv record, formed by concatenating the arXiv paper's authors and title, using the prompt in Table \ref{tab:prompt_check_match}. Only entries that pass this verification step are considered to have a valid arXiv version. Finally, we validate external references by checking that the cited theorem identifier appears in the known list of theorems extracted from the matched arXiv paper. This process yields tuples of the form (citing arXiv ID, citing theorem ID, cited arXiv ID, cited theorem ID), which define the theorem-level citation edges.

The key advantage of using LLMs for theorem-level citation extraction is their ability to leverage natural language understanding to identify the theorems that are genuinely used in a proof. This stands in contrast to rule-based approaches that rely on regular expression matching, which may incorrectly treat purely comparative mentions as logical dependencies, or struggle to associate proofs deferred to later sections with the corresponding theorem statements. Such rule-based methods often require extensive ad hoc heuristics, whereas the LLM-based approach provides a more robust and semantically faithful extraction of theorem dependencies.

To construct conjecture-level edges, we first embed each conjecture into a vector representation and compute pairwise cosine similarities. For each conjecture, we retain the top 30 most similar conjectures according to cosine similarity. We then employ DeepSeek-V3.2 to classify each conjecture pair into one of four relation types: same, implication (one conjecture implies the other), related, or unrelated.

Our final graph contains 6,681 conjecture nodes, 102,185 paper nodes, 1,880,586 theorem nodes and 64,563 author nodes.

\begin{table}[t]
\caption{Prompt for extracting the title of a bibliographic item.}
\label{tab:prompt_extract_title}
\centering
\begin{tabular}{p{0.9\linewidth}}
\toprule
\ttfamily
You are a helpful assistant. Please extract the title from the following bibliographic entry and output the title in JSON format as follows: \{"title": "the title you extracted"\}.\\
\ttfamily
\{bibliographic entry\}\\
\bottomrule
\end{tabular}
\end{table}

\begin{table}[t]
\caption{Prompt for checking whether a bibliographic entry matches an arXiv item.}
\label{tab:prompt_check_match}
\centering
\begin{tabular}{p{0.9\linewidth}}
\toprule
\ttfamily
You are an expert in mathematics. You will be given a bibliographic entry of a published paper and a bibliographic entry of an arXiv preprint. Your task is to determine if the published paper is the final version of the arXiv preprint by comparing the authors and titles. Note that an author's name may appear in different variants, such as full names or abbreviations. Additionally, some authors' names may be followed by their institutional affiliations. If the published paper is the final version of the arXiv preprint, please answer "Yes." If it is not, please answer "No."\\
\ttfamily
Now, please evaluate the following published paper and arXiv preprint:\\
\ttfamily
**Published Paper:** \{bibliographic entry\}\\
\ttfamily
**Arxiv Preprint:** \{arXiv bibliographic entry\}\\
\bottomrule
\end{tabular}
\end{table}

\section{List of Journals}\label{sec:app_journals}
Tables \ref{tab:journals_0} and \ref{tab:journals_1} list the journals included in our study.

\begin{table}[h]
\caption{List of journals we use. Journal names follow the MathSciNet format, which may differ slightly from common usage. For example, Ann. of Math. (2) refers to Ann. of Math.}\label{tab:journals_0}
\centering
\resizebox*{!}{\dimexpr\textheight-3\baselineskip\relax}{%
\begin{tabular}{ll}
\toprule
ACM Trans. Algorithms & Discrete Anal. \\
ALEA Lat. Am. J. Probab. Math. Stat. & Discrete Comput. Geom. \\
Acta Arith. & Discrete Contin. Dyn. Syst. \\
Acta Math. & Discrete Math. \\
Acta Numer. & Doc. Math. \\
Adv. Math. & Duke Math. J. \\
Adv. Theor. Math. Phys. & ESAIM Control Optim. Calc. Var. \\
Adv. in Appl. Math. & ESAIM Math. Model. Numer. Anal. \\
Algebr. Geom. & Electron. Commun. Probab. \\
Algebr. Geom. Topol. & Electron. J. Combin. \\
Algebra Number Theory & Electron. J. Probab. \\
Amer. J. Math. & Electron. J. Stat. \\
Anal. PDE & Ergodic Theory Dynam. Systems \\
Ann. Appl. Probab. & European J. Combin. \\
Ann. Henri Poincaré & Exp. Math. \\
Ann. Inst. Fourier (Grenoble) & Forum Math. \\
Ann. Inst. Henri Poincaré Probab. Stat. & Forum Math. Pi \\
Ann. Probab. & Forum Math. Sigma \\
Ann. Pure Appl. Logic & Found. Comput. Math. \\
Ann. Sc. Norm. Super. Pisa Cl. Sci. (5) & Fund. Math. \\
Ann. Sci. Éc. Norm. Supér. (4) & Geom. Dedicata \\
Ann. Statist. & Geom. Funct. Anal. \\
Ann. of Math. (2) & Geom. Topol. \\
Appl. Comput. Harmon. Anal. & Groups Geom. Dyn. \\
Arch. Ration. Mech. Anal. & Homology Homotopy Appl. \\
Asian J. Math. & IEEE Trans. Automat. Control \\
Astérisque & IEEE Trans. Inform. Theory \\
Atti Accad. Naz. Lincei Rend. Lincei Mat. Appl. & Illinois J. Math. \\
Bernoulli & Indiana Univ. Math. J. \\
Bull. Amer. Math. Soc. (N.S.) & Int. Math. Res. Not. IMRN \\
Bull. Lond. Math. Soc. & Internat. J. Algebra Comput. \\
Bull. Soc. Math. France & Internat. J. Math. \\
C. R. Math. Acad. Sci. Paris & Invent. Math. \\
Calc. Var. Partial Differential Equations & Inverse Probl. Imaging \\
Camb. J. Math. & Inverse Problems \\
Canad. J. Math. & Israel J. Math. \\
Combin. Probab. Comput. & J. ACM \\
Combinatorica & J. Algebra \\
Comm. Anal. Geom. & J. Algebraic Combin. \\
Comm. Partial Differential Equations & J. Algebraic Geom. \\
Comm. Pure Appl. Math. & J. Amer. Math. Soc. \\
Comment. Math. Helv. & J. Anal. Math. \\
Commun. Comput. Phys. & J. Combin. Theory Ser. A \\
Commun. Contemp. Math. & J. Combin. Theory Ser. B \\
Commun. Math. & J. Comput. Phys. \\
Commun. Math. Sci. & J. Differential Equations \\
Commun. Number Theory Phys. & J. Differential Geom. \\
Compos. Math. & J. Eur. Math. Soc. (JEMS) \\
Comput. Complexity & J. Funct. Anal. \\
Comput. Methods Appl. Mech. Engrg. & J. Geom. Anal. \\
\bottomrule
\end{tabular}
}
\end{table}

\begin{table}[h]
\caption{List of journals we use (continued).}\label{tab:journals_1}
\centering
\resizebox{\linewidth}{!}{%
\begin{tabular}{ll}
\toprule
J. Geom. Phys. & Pacific J. Math. \\
J. High Energy Phys. & Phys. D \\
J. Inst. Math. Jussieu & Phys. Rev. A \\
J. Lond. Math. Soc. & Phys. Rev. E \\
J. Mach. Learn. Res. & Phys. Rev. Lett. \\
J. Math. Anal. Appl. & Probab. Surv. \\
J. Math. Phys. & Probab. Theory Related Fields \\
J. Math. Pures Appl. & Proc. Amer. Math. Soc. \\
J. Mod. Dyn. & Proc. Lond. Math. Soc. \\
J. Number Theory & Proc. Natl. Acad. Sci. USA \\
J. Phys. A & Publ. Math. Inst. Hautes Études Sci. \\
J. Pure Appl. Algebra & Publ. Res. Inst. Math. Sci. \\
J. Reine Angew. Math. & Pure Appl. Math. Q. \\
J. Sci. Comput. & Q. J. Math. \\
J. Stat. Phys. & Quantum Topol. \\
J. Symbolic Logic & Random Matrices Theory Appl. \\
J. Symplectic Geom. & Random Structures Algorithms \\
J. Topol. & Represent. Theory \\
Jpn. J. Math. & Res. Math. Sci. \\
Kinet. Relat. Models & Rev. Mat. Iberoam. \\
Kyoto J. Math. & SIAM J. Appl. Dyn. Syst. \\
Lett. Math. Phys. & SIAM J. Appl. Math. \\
Manuscripta Math. & SIAM J. Comput. \\
Math. Ann. & SIAM J. Control Optim. \\
Math. Comp. & SIAM J. Discrete Math. \\
Math. Models Methods Appl. Sci. & SIAM J. Imaging Sci. \\
Math. Oper. Res. & SIAM J. Math. Anal. \\
Math. Proc. Cambridge Philos. Soc. & SIAM J. Numer. Anal. \\
Math. Program. & SIAM J. Optim. \\
Math. Res. Lett. & SIAM J. Sci. Comput. \\
Math. Z. & SIAM Rev. \\
Mathematika & SIGMA Symmetry Integrability Geom. Methods Appl. \\
Mem. Amer. Math. Soc. & Sci. China Math. \\
Michigan Math. J. & Selecta Math. (N.S.) \\
Mosc. Math. J. & Stochastic Process. Appl. \\
Multiscale Model. Simul. & Theory Comput. \\
Nagoya Math. J. & Topology Appl. \\
Nonlinear Anal. & Tr. Mat. Inst. Steklova \\
Nonlinearity & Trans. Amer. Math. Soc. \\
Notices Amer. Math. Soc. & Transform. Groups \\
\bottomrule
\end{tabular}
}
\end{table}

\end{document}